\documentstyle[12pt]{article}

\newcommand {\nn}    {\nonumber}
\newcommand {\vs}[1]  { \vspace*{#1 cm} }

\newcounter{eq}
\newcounter{sc}

\newcommand {\NP}   {Nucl.Phys.}
\newcommand {\PL}   {Phys.Lett.}
\newcommand {\PR}   {Phys.Rev.}
\newcommand {\PRL}   {Phys.Rev.Lett.}

\newcommand {\AP}   {Ann.of Phys.}

\newcommand {\JHEP}  {J.High Energy Phys.}

\def\overleftrightarrow#1{\vbox{\ialign{##\crcr
 $\leftrightarrow$\crcr\noalign{\kern-1pt\nointerlineskip}
 $\hfil\displaystyle{#1}\hfil$\crcr}}}

\newlength{\minitwocolumn}
\begin{document}

\begin{flushright}
CHIBA-EP-111\\
EDO-EP-25\\
November, 1998\\
\end{flushright}
\vspace{20pt}

\pagestyle{empty}
\baselineskip15pt

\begin{center}
{\large\bf Duality of Super D-brane Actions \\
in \\
Type II Supergravity Background
 \vskip 1mm
}

\vspace{10mm}

Tadahiko Kimura
          \footnote{
          E-mail address:\ kimura@cuphd.nd.chiba-u.ac.jp
                  }
\\

\vspace{5mm}
          Department of Physics, Faculty of Science,
          Chiba University, Chiba 263, JAPAN \\

\vspace{5mm}

and

\vspace{5mm}

Ichiro Oda
          \footnote{
          E-mail address:\ ioda@edogawa-u.ac.jp
                  }
\\
\vspace{5mm}
          Edogawa University,
          474 Komaki, Nagareyama City, Chiba 270-0198, JAPAN \\

\end{center}


\vspace{5mm}
\begin{abstract}
We show that supersymmetric and $\kappa$-symmetric Dp-brane actions
in type II supergravity background have the same duality transformation
properties as those in a flat Minkowskian background.
Specially, it is shown that the super D-string transforms in a
covariant way while the super D3-brane is self-dual under the $SL(2,Z)$
duality. Also, the D2-brane and the D4-brane transform in ways
expected from the relation between type IIA superstring theory and
M-theory. The present study proves that various duality symmetries,
which were originally found in the flat background field, are precisely
valid even in the curved background geometry.
\vspace{3mm}
\begin{flushleft}
{\it PACS:} 11-25.-w; 12.60.Jv; 04.65 \\
{\it Key words:} D-branes; Duality; Supergravity
\end{flushleft}

\end{abstract}

\newpage
\pagestyle{plain}
\pagenumbering{arabic}


\rm
\section{Introduction}

One of the most important insights about nonperturbative
behavior of superstring theory \cite{GSW} and M-theory 
\cite{Town, Witten2} is the existence of the intricate network 
of duality symmetries.  Much nonperturbative
information has been gleaned by exploring the duality transformations
connecting various string vacua and excitations.
For instance, it was remarkably shown that the five known superstring
theories and the still rather mysterious M-theory are indeed
nonperturbatively equivalent by means of the web of string
dualities \cite{Schwarz0}. This is a typical example that a chain
of duality transformations reduces the degree of non-uniqueness
of string theory. Another interesting aspect of the duality
transformations is that it allows us to study a strongly coupled
string theory by mapping that theory to a weakly coupled dual
theory whenever such a dual theory exists. 

Among three types of dualities appearing in superstring theory,
named S-, T- and U-dualities, S-duality possesses a curious
historical position. The appearance of a non-compact global
symmetry group $G$ was already known in the 1970's to be a
characteristic feature of supergravity theory, which originally
arose from the desire to include supersymmetry into the
framework of general relativity with the hope that supersymmetry
might tame the notorious ultraviolet divergences but is now
considered to be the low energy effective theory of superstring theory.
Such a global group is realized nonlinearly by the scalar fields
parametrizing the coset space $G/H$ where $H$ is the maximal
compact subgroup of $G$ \cite{Cremmer1, Cremmer2, Salam}.
However, in those days, the non-compact global symmetry
was regarded as an artifact of the low energy approximation
to underlying renormalizable theory and not taken seriously.
It was recognized only recently that the $SL(2,Z)$ subgroup
of the $SL(2,R)$ symmetry of Cremmer et al. \cite{Cremmer1}
precisely corresponds to one example of S-duality in the
toroidally compactified heterotic superstring.

Another line of development is the discovery of super D-branes carrying
Ramond-Ramond charge \cite{Polchinski}. These extended objects
have been shown to play a fundamental role in nonperturbative
aspects of string theory and quantum black holes.
It is interesting to notice that historically the
D-branes were first studied using T-duality symmetry on the string
world-sheet \cite{Polchinski}, and their role in the web of string
dualities is at present understood to some extent.

Despite these impressive developments, our understanding about
nonperturbative regime in superstring theory is still far
from complete. Perhaps, under such a situation, a sound direction
of study is to achieve as thorough an understanding of symmetry
and dynamics of super D-branes as possible in order to clarify
nonpertubative aspects of superstring theory.
Actually motivated by this reason and more, in recent years,
the world-volume actions for supersymmetric and
$\kappa$-symmetric D-branes have been constructed
in a flat background \cite{Aganagic1}
as well as a general type II supergravity background
\cite{Cederwall1, Cederwall2, Bergshoeff}, and some of
duality symmetries have been investigated.

One of the main distinctions between super Dp-branes and
super p-branes is that in the former there is an abelian gauge
field $A_i$ in addition to the superspace coordinates
$(X^i, \theta)$. Then by carrying out a duality transformation of
this abelian gauge potential, we can arrive at dual super D-brane
actions. In the case of a flat background, it has been already
observed that the resulting world-volume actions after a duality
transformation give rise to the expected properties \cite{Aganagic2}.
Specifically, it was shown that
the super D-string transforms in a covariant manner while the super
D3-brane is self-dual under the $SL(2,Z)$ S-duality. Also, the D2-brane
and the D4-brane transform in ways expected from the duality relation
between type IIA superstring theory and M-theory.

Until recently, however, we have had no idea whether or not these
dualities symmetries existing in the (super-) D-brane actions in
the flat background also exist in a curved background.
A first step towards the proof of the $SL(2,Z)$ S-duality of the
super D-string and D3-brane actions on $AdS_5 \times S^5$ was
taken in \cite{Oda1, Oda2} where it was shown that this duality
indeed exists in the specific background, and later one of the
present authors has also verified this fact without fixing
$\kappa$-symmetry \cite{Kimura}. The main motivation of
the present study is to prove that various duality symmetries found
in the flat background field, are valid even in a more general curved
background geometry.

A preliminary work of a super D-string in type IIB
supergravity background was reported by one of authors \cite{Oda0}
where it was shown that the super D-string in this curved background
is transformed to the type IIB Green-Schwarz superstring action
\cite{GS},
thereby proving the $SL(2,Z)$ covariance of the super D-string.
In this paper, as promised in \cite{Oda0}, we will extend the ideas
to broader situations, those are, the self-duality of the super
D3-brane in type IIB supergravity background under an $SL(2,Z)$
S-duality transformation, and the relations between D2 and M2-branes
and the one between D4 and M5-branes in type IIA supergravity
background which are expected from the IIA/M-duality.

This article is organized as follows.
Section 2 reviews super D-brane actions in a
general IIA and IIB supergravity background \cite{Cederwall2,
Bergshoeff}.
In Section 3 it is then proved in a quantum-mechanically
exact manner that the super D-string action in type IIB on-shell
supergravity background is transformed to the type IIB Green-Schwarz
superstring action with the $SL(2,Z)$ covariant tension
through an S-duality transformation.
Section 4 deals with the super D2-brane  in type IIA on-shell
supergravity background and presents that the super D2-brane action
can be transformed to the super M2-brane action with a circular
eleventh dimension by a duality transformation.
In Section 5 we show that the super D3-brane action
in type IIB on-shell supergravity background is mapped into
itself by an S-duality transformation,
thereby verifying the $SL(2,Z)$ self-duality of the action.
We shall present both
classical and quantum-mechanical proofs here.
In Section 6 it is shown that the super D4-brane
action  becomes identical to the supersymmetric action
which is obtained in terms of
double-dimensional reduction of the super M5-brane
action in the eleven dimensional space-time through
a duality transformation.
The final section will be devoted to discussions.

To close this section, we would like to stress that the analysis
considered in this paper has two improvements over that in a
paper \cite{Aganagic2} even if there exists an exact correspondence
of the obtained results. One big improvement, of course, Aganagic
et al. \cite{Aganagic2} have taken account of only a flat
background while we have
considered a general curved background. Another important improvement
is that their analysis is purely classical, on the other hand, we
have performed the quantum analysis at least for the super D-string
and D3-brane because of their importance.

\section{ Super D-brane actions in a general type II background}

We start by reviewing super Dp-brane actions in a general
type II supergravity background \cite{Cederwall2, Bergshoeff}.
It is well known nowadays that super Dp-brane actions consist
of two terms, those are, the Dirac-Born-Infeld action and
the Wess-Zumino action. The former includes the NS-NS two-form,
dilaton and world-volume metric in addition to Abelian gauge
field while the latter
action contains the coupling of the D-brane to the R-R fields.
The two terms are separately invariant under type II superspace
reparametrizations as well as $(p+1)$-dimensional general
coordinate transformations. However, local $\kappa$ symmetry
is achieved by a suitable conspiracy between the two terms.

Then super Dp-brane actions in a general type II on-shell
supergravity background which we consider are given by
\begin{eqnarray}
S = S_{DBI} + S_{WZ},
\label{2.1}
\end{eqnarray}
with
\begin{eqnarray}
&{}& S_{DBI} = -  \int_{M_{p+1}} d^{p+1} \sigma
\sqrt{- \det ( G_{ij} + {\cal F}_{ij} )}, \nn\\
&{}& S_{WZ} =  \int_{M_{p+1}} e^{\cal{F}}
\wedge C = \int_{M_{p+1}} \Omega_{p+1}
= \int_{M_{p+2}} I_{p+2},
\label{2.2}
\end{eqnarray}
where $\sigma^i \ (i = 0, 1, \ldots, p)$ are the world-volume
coordinates, and $G_{ij}$ is the metric of 
the world-volume.
We have defined various quantities as follows:
\begin{eqnarray}
{\cal F} &=& F - b_2, \nn\\
F &=& dA, \nn\\
C &=& \displaystyle \bigoplus_{n=0}^{9} C_{(n)}, \nn\\
I_{p+2} &=& d \Omega_{p+1} = d ( e^{\cal{F}} \wedge C ), \nn\\
M_{p+1} &=& \partial M_{p+2},
\label{2.3}
\end{eqnarray}
where $F$ is the Maxwell field strength 2-form, and the 2-form
$b_2$ is introduced such that ${\cal{F}}$ is invariant under
supersymmetry. And the RR $n$-form fields $C_{(n)}$ are collected
in $C$ with $n$ taking odd integers for type IIA and even integers
for type IIB.

In addition, in order to describe the curved target superspace geometry
we have to introduce the superspace vielbein 1-form $E^A$ defined by
\begin{eqnarray}
E^A = dZ^M E_M^A,
\label{2.4}
\end{eqnarray}
with $dZ^M$ denoting the superspace differential $(dX^m, d\theta^\mu)$,
and the torsion 2-form $T^A = DE^A$ as well as the curvature
2-form defined in terms of the spin connection $\omega_A^B$ as
\begin{eqnarray}
R_A^B = d\omega_A^B + \omega_A^C \wedge \omega_C^B.
\label{2.5}
\end{eqnarray}
Note that we have also defined as $M = (m, \mu)$ in curved superspace
while $A = (a, \alpha)$ in flat superspace as usual.
Then the world-volume metric
$G_{ij}$ is represented by
\begin{eqnarray}
G_{ij} = E_i^a E_j^b \eta_{ab},
\label{2.6}
\end{eqnarray}
where $E_i^A = \partial_i Z^M E_M^A$ and $\eta_{ab}$
= diag$(-,+, \ldots, +)$.

Throughout this paper we use following conventions for superspace
forms. Firstly, a general $n$-form superfield $\Omega_{(n)}$
is expanded as
\begin{eqnarray}
\Omega_{(n)} &=& \frac{1}{n!} dZ^{M_n} \wedge \ldots \wedge dZ^{M_1}
\Omega_{M_1 \ldots M_n}, \nn\\
&=& \frac{1}{n!} E^{A_n} \wedge \ldots \wedge E^{A_1}
\Omega_{A_1 \ldots A_n}.
\label{2.7}
\end{eqnarray}
Secondly, we define the exterior derivative as an operator acting
from the right
\begin{eqnarray}
d (\Omega_{(m)} \wedge \Omega_{(n)}) = \Omega_{(m)} \wedge d\Omega_{(n)}
+ (-)^n d\Omega_{(m)} \wedge \Omega_{(n)}.
\label{2.8}
\end{eqnarray}

Now, following the paper \cite {Cederwall2}, let us define the NS-NS
3-form superfield $H_3$ and the R-R $n$-form superfield
$R$ as \footnote{See the ref.\cite{Bergshoeff}
for type IIA massive supergravity, i.e., $R_{0}=m$}
\begin{eqnarray}
H_{3} &=& db_2, \nn\\
R &=& e^{b_2} \wedge d( e^{-b_2} \wedge C )
= \displaystyle \bigoplus_{n=1}^{10} R_{(n)}.
\label{2.9}
\end{eqnarray}
It is obvious that from these definitions the field strengths
obey the following Bianchi identities
\begin{eqnarray}
dH_{3} &=& 0, \nn\\
e^{b_2} \wedge d( e^{-b_2} \wedge R )
= dR - R \wedge H_{3} &=& 0.
\label{2.10}
\end{eqnarray}

In order to reduce the enormous unconstrained field content
included in the superfields to the field content of the on-shell
type II supergravity theory, one has to impose the constraints
on the torsion and the field strengths by hand, which make various
Bianchi identities to coincide with the equations of motion of
supergravity.
Under the assumption of vanishing
(or constant) dilaton, the nontrivial constraints imposed on the
torsion and field strength components \cite{Cederwall2} take the
following forms for
type IIA:
\begin{eqnarray}
T_{\alpha\beta}^c &=& 2i \gamma_{\alpha\beta}^c, \nn\\
H_{a \alpha\beta} &=& - 2i (\gamma_{11} \gamma_a)_{\alpha\beta}, \nn\\
R_{(n) a_1 \ldots a_{n-2} \alpha\beta} &=&  2i
(\gamma_{a_1 \ldots a_{n-2}} (\gamma_{11})^{\frac{n}{2}})_{\alpha\beta},
\label{2.11}
\end{eqnarray}
and for IIB:
\begin{eqnarray}
T_{\alpha\beta}^c &=& 2i \gamma_{\alpha\beta}^c, \nn\\
H_{a \alpha\beta} &=& - 2i ({\cal K} \gamma_a)_{\alpha\beta}, \nn\\
R_{(n) a_1 \ldots a_{n-2} \alpha\beta} &=&  2i
(\gamma_{a_1 \ldots a_{n-2}} {\cal K}^{\frac{n-1}{2}}
{\cal E})_{\alpha\beta},
\label{2.12}
\end{eqnarray}
where ${\cal E}$, ${\cal I}$, and ${\cal K}$
describing the $SO(2)$ matrices are defined in terms of the
conventional Pauli matrices $\sigma_i$ as follows:
\begin{eqnarray}
{\cal E} = i \sigma_2 = \pmatrix{
0  & 1 \cr -1 & 0 \cr }, \ {\cal I} = \sigma_1 = \pmatrix{
0  & 1 \cr 1 & 0 \cr },  \ {\cal K} = \sigma_3 = \pmatrix{
1  & 0 \cr 0 & -1 \cr }.
\label{2.13}
\end{eqnarray}

Based on this formulation of the super Dp-brane actions in type
II on-shell supergravity background, we shall explore various
duality symmetries in subsequent sections. 
Of course, our formulation is not so
general in that we have confined ourselves to the vanishing
(or constant) dilaton and antisymmetric tensor fields e.t.c.
It is quite valuable to remove these restrictions and construct
a more general formalism in future.

\section{The super D-string}

In this section we would like to consider the super D-string
(i.e. the super D1-brane) first. The super D2, D3 and D4-branes
will be treated in order in subsequent sections. In these sections
we shall prove various duality symmetries of the super D-brane
actions in type II on-shell supergravity background. The corresponding
proofs have been already done in ref.\cite{Aganagic2} in the case of a
flat Minkowskian background. Actually, we will see that the duality
relations found in ref.\cite{Aganagic2} precisely hold even in
the curved background. This fact is quite important for future
development in string theory and M-theory since the global
discrete symmetries such as the $SL(2,Z)$ S-duality are nowadays
believed to be exact symmetries in still mysterious underlying theory
\cite{Hull, Witten2} so that these symmetries should be valid in arbitrary
curved
background geometries.

In the case at hand, the action (\ref{2.1}), (\ref{2.2}) and the
constraints (\ref{2.12}) reduce to
\begin{eqnarray}
S &=& S_{DBI} + S_{WZ}, \nn\\
S_{DBI} &=& - \int_{M_2} d^2 \sigma
\sqrt{- \det ( G_{ij} + {\cal F}_{ij} )}, \nn\\
S_{WZ} &=& \int_{M_2 = \partial M_3} C_2 = \int_{M_3}
I_3, \nn\\
H_{3} &=& db_2 = i \bar{E} \wedge \hat{E} \wedge {\cal K} E, \nn\\
I_3 &=& dC_2 = - i \bar{E} \wedge \hat{E} \wedge {\cal I} E,
\label{3.1}
\end{eqnarray}
where we have used not only the superspace convention (\ref{2.7})
but also the fact that the R-R 3-form field strength superfield
$R_{(3)}$ coincides with the Wess-Zumino form $I_3$. Moreover,
for a while we have neglected the axion 
$C_{(0)}$ which will be considered later. 
In Eq.(\ref{3.1}), $\bar{E}$, $\hat{E}$ and $E$
represent the Dirac conjugate of $E^{I\alpha}$, $E^a \gamma_a$
and $E^{I\alpha}$ with $I$ being the $N=2$ index ($I=1, 2$),
respectively.

Now we are ready to present a quantum-mechanical exact proof
of $SL(2,Z)$ S-duality covariance of the super D-string action
in a general ten dimensional IIB supergravity background
\cite{Oda0}.
To this end, the crucial observations concern the fact that the action
in Eq.(\ref{3.1}) is of the form similar to that on
$AdS_5 \times S^5$ \cite{Oda1}. Once this point is understood,
the analysis is a fairly straightforward generalization of that
presented in \cite{Oda1}, though some points are a little more
involved. Following the techniques developed in \cite{de Alwis, Oda4},
let us utilize the path integral of the first-order Hamiltonian form.

As the first step of the Hamiltonian formalism, let us introduce
the canonical conjugate momenta $\pi^i$ corresponding to the
gauge field $A_i$ defined as
\begin{eqnarray}
\pi^i \equiv \frac{\partial S}
{\partial \dot{A}_i} =  \frac{\partial S_{DBI}}
{\partial \dot{A}_i},
\label{3.3}
\end{eqnarray}
where we used the fact that the Wess-Zumino term is independent
of the gauge potential, which holds only in the case of string
theory. Then the canonical conjugate momenta $\pi^i$ are calculated to
be
\begin{eqnarray}
\pi^0 = 0, \ \pi^1 = \frac{{\cal F}_{01}}
{\sqrt{-\det ( G_{ij} + {\cal F}_{ij} )}},
\label{3.4}
\end{eqnarray}
where the former equation just shows the existence of the $U(1)$
gauge invariance.  From these equations we will see that the
Hamiltonian density takes the form
\begin{eqnarray}
{\cal H} = \sqrt{1 + (\pi^1)^2} \sqrt{- \det G_{ij}}
- A_0 \partial_1 \pi^1 + \partial_1 (A_0 \pi^1)
+ \pi^1 b_{01} - C_{01},
\label{3.5}
\end{eqnarray}

Now the partition function is defined by the first-order
Hamiltonian form with respect to only the gauge field as follows:
\begin{eqnarray}
Z &=& \frac{1}{\int {\cal D}\pi^0} \int {\cal D}\pi^0
{\cal D}\pi^1 {\cal D}A_0 {\cal D}A_1
\exp{ i \int d^2 \sigma ( \pi^1 \partial_0 A_1 - {\cal H} ) } \nn\\
&=& \int {\cal D}\pi^1 {\cal D}A_0 {\cal D}A_1
\exp{ i \int d^2 \sigma} \nn\\
& & {} \times \left[ - A_1 \partial_0 \pi^1 + A_0 \partial_1
\pi^1 -  \sqrt{1 + (\pi^1)^2} \sqrt{- \det G_{ij}}
- \pi^1 b_{01} + C_{01} - \partial_1 ( A_0 \pi^1 ) \right].
\label{3.6}
\end{eqnarray}

Provided that we take the boundary conditions for $A_0$ such that
the last surface term in the exponential identically vanishes,
then we can carry out the integrations over $A_i$ explicitly,
which gives rise to $\delta$ functions
\begin{eqnarray}
Z &=& \int {\cal D}\pi^1 \delta(\partial_0 \pi^1)
\delta(\partial_1 \pi^1) \exp{ i \int d^2 \sigma
\left[ -\sqrt{1 + (\pi^1)^2} \sqrt{- \det G_{ij}}
+ C_{01} - \pi^1 b_{01} \right] }.
\label{3.7}
\end{eqnarray}
The existence of the $\delta$ functions reduces the integral
over $\pi^1$ to the one over only its zero-modes. If we require
that one space component is compactified on a circle, these
zero-modes are quantized to be integers \cite{Witten}.
As a consequence, the partition function becomes
\begin{eqnarray}
Z &=& \displaystyle{ \sum_{m \in {\bf Z}} } \exp{ i \int d^2 \sigma
\left[ -\sqrt{1 + m^2} \sqrt{- \det G_{ij}}
+ C_{01} - m b_{01} \right] },
\label{3.8}
\end{eqnarray}
from which we can read off the effective action
\begin{eqnarray}
S = \int d^2 \sigma \left( -\sqrt{1 + m^2} \sqrt{- \det
G_{ij}} + C_{01} - m b_{01} \right).
\label{3.9}
\end{eqnarray}
Moreover, recalling the relations in (\ref{3.1})
\begin{eqnarray}
\int_{M_2 = \partial M_3} d^2 \sigma
( C_{01} - m b_{01} ) =  \int_{M_3}
( I_3 - m H_3 ) = -i \int_{M_3}
\bar{E} \wedge \hat{E} \wedge ( m {\cal K} + {\cal I} ) E,
\label{3.10}
\end{eqnarray}
and then carrying out an orthogonal transformation
\begin{eqnarray}
U^T (m {\cal K} + {\cal I}) U = - \sqrt{1 + m^2} {\cal K},
\label{3.11}
\end{eqnarray}
with an orthogonal matrix
$U = \frac{1}{\sqrt{1 + (m - \sqrt{1+m^2})^2}}[(m - \sqrt{1+m^2})1
- {\cal E}]$, one finally obtains the action
\begin{eqnarray}
S = -\sqrt{1 + m^2} \left( \int_{M_2} d^2 \sigma   \sqrt{- \det
G_{ij}}  -i \int_{M_3} \bar{E} \wedge \hat{E} \wedge {\cal K} E
\right).
\label{3.12}
\end{eqnarray}
This is nothing but type IIB Green-Schwarz superstring action
with the modified tension $\sqrt{1 + m^2}$ in a type IIB
supergravity background \cite{Grisaru}.

It is worthwhile to notice that the result obtained above agrees with
the tension formula for the $SL(2,Z)$ S-duality spectrum
of strings in the flat background \cite{Schwarz} provided
that we identify the integer value $\pi^1 = m$ with the 
NS-NS charge corresponding to the $(m,1)$ string.
This identification means that the D-string action is actually
the action for an arbitrary number of 'fundamental' IIB strings
bound to a single D-string.
To show more clearly that the tension
at hand is the $SL(2,Z)$ covariant tension, it
would be more convenient to start with the following classical action
\begin{eqnarray}
S = -  n \int_{M_2} d^2 \sigma
\left[ e^{-\phi} \bigl( \sqrt{- \det ( G_{ij} + {\cal F}_{ij} )}
- C_2 \bigr)
+ \frac{1}{2} \epsilon^{ij} \chi F_{ij} \right],
\label{3.13}
\end{eqnarray}
where $n$ is an integer, and we have introduced the constant
dilaton $\phi$ and the constant axion $C_{(0)} \equiv \chi$.
Then following the same path
of derivation as above, we can obtain the manifestly $SL(2,Z)$
covariant tension
\begin{eqnarray}
T = \sqrt{ (m + n \chi)^2 +  n^2 e^{-2\phi}}.
\label{3.14}
\end{eqnarray}

Here we would like to comment two important points. One point is
that we have shown that there exists $SL(2,Z)$ S-duality in type
IIB superstring theory even in a general type IIB supergravity
background without reference to any approximation. Thus this relation
is quantum-mechanically exact.

The other point is the problem of whether one can interpret the
orthogonal transformation (\ref{3.11}) as the $SO(2)$ rotation of
the $N=2$ spinor coordinates.
In our previous paper \cite{Oda1, Oda2} this problem
was emphasized too much, but on reflection it turns out that this
problem is rather trivial by the following reasons.
Notice that the torsion constraint in (\ref{2.10}) is obviously
invariant under this rotation. Moreover,
since we require that the original super D-string action and the
fundamental Green-Schwarz action reduce to the well-known forms
of the corresponding flat space actions in the flat space limit,
$E^{I\alpha}$ with the $SO(2)$ index $I$ and $E^a$ must take the
following forms at the lowest order expansion with respect to
the spinor coordinates $\theta$
\begin{eqnarray}
E^I_i &=& \partial_i \theta^I + \ldots, \nn\\
E^a_i &=&  \partial_i X^a - i \bar{\theta}^I \gamma^a \partial_i
\theta^I + \ldots,
\label{3.15}
\end{eqnarray}
where the dots indicate the higher order terms reflecting the
curved nature of the background metric.
These facts mean that $E^I$ transforms as the adjoint representation
of the $SO(2)$ group, on the other hand, $E^a$ must be invariant
under an $SO(2)$ rotation. Accordingly, we
can understand that the orthogonal transformation (\ref{3.11})
is indeed performed by an $SO(2)$ rotation of the $N=2$ spinor
coordinates.
In this way, we have succeeded in deriving the $SL(2,Z)$
S-duality of type IIB superstring theory in type IIB on-shell
supergravity background at least within the
present context.

\section{The super D2-brane}
Next we turn to the $\it{classical}$ derivation of a duality
transformation between the super D2-brane
(i.e., the super D-membrane) in type IIA
supergravity background and the super M2-brane in eleven dimensional
supergravity. The authors in a paper \cite{Bergshoeff} have already
dealt with this problem from a different viewpoint. The method adopted
there is start with the super M2-brane in eleven dimensions, achieve
the dimensional reduction to ten dimensions a la KK ansatz, then
perform a duality transformation for the purpose of getting the
super D2-brane action and its $\kappa$-symmetry. Our method is
similar to that of Aganagic et al. \cite{Aganagic2} where
the above arguments were reversed, namely, the super M2-brane action
was obtained from starting with
the super D2-brane action through a duality transformation
of the world-volume gauge field.

{}From Eqs.(\ref{2.1}) and (\ref{2.2}), the super D2-brane action in
the string metric becomes
\begin{eqnarray}
S &=& S_{DBI} + S_{WZ} + S_{\tilde{H}}, \nn\\
S_{DBI} &=& - \int_{M_3} d^3 \sigma
\sqrt{- \det ( G_{ij} + {\cal F}_{ij} )}, \nn\\
S_{WZ} &=& \int_{M_3 = \partial M_4} ( C_3 + C_1 \wedge
{\cal F} ) = \int_{M_4} I_4, \nn\\
S_{\tilde{H}} &=& \int_{M_3} d^3 \sigma \frac{1}{2}
\tilde{H}^{ij} ( F_{ij} - 2 \partial_i A_j ),
\label{4.1}
\end{eqnarray}
where we have added $S_{\tilde{H}}$ to the original action to
perform a duality transformation. Moreover, in this case
the constraints (\ref{2.11}) on the field strengths reduce to
\begin{eqnarray}
H_3 &=& db_2 = i \bar{E} \wedge \gamma_{11} \hat{E} \wedge E, \nn\\
R_{(4)} &=& \frac{i}{2} \bar{E} \wedge \gamma_{ab} E \wedge
E^b \wedge E^a, \nn\\
R_{(2)} &=& i \bar{E} \wedge \gamma_{11} E.
\label{4.2}
\end{eqnarray}
{}From these equations and the definitions (\ref{2.3}) and (\ref{2.9}),
we find that $C_3$ and $C_1$ are determined by the conditions
\begin{eqnarray}
R_{(4)} &=& dC_3 + db_2 \wedge C_1 =
\frac{i}{2} \bar{E} \wedge \gamma_{ab} E \wedge E^b \wedge E^a, \nn\\
R_{(2)} &=& dC_1 = i \bar{E} \wedge \gamma_{11} E.
\label{4.3}
\end{eqnarray}

At this stage, we take the variation with respect to $A_i$,
which gives us the solution $\tilde{H}^{ij}
= \epsilon^{ijk} \partial_k B$ with $B$ being a scalar superfield.
Then after substituting this solution into the action and solving
the equation of motion for $F_{ij}$ in order to rewrite the action
in terms of $B$ instead of $F_{ij}$, we arrive at the dual action
$S_D$ of (\ref{4.1})
\begin{eqnarray}
S_D = - \int_{M_3} d^3 \sigma \sqrt{- \det G'_{ij}}
+ \int_{M_3} ( C_3 + b_2 \wedge dB ),
\label{4.4}
\end{eqnarray}
where we have defined as
\begin{eqnarray}
G'_{ij} = G_{ij} + (\partial_i B + C_i)(\partial_j B + C_j).
\label{4.5}
\end{eqnarray}
Incidentally, in order to derive the dual action we have used
the mathematical formulas holding for $3 \times 3$ matrices
\begin{eqnarray}
\det ( G_{ij} + A_i A_j ) &=& (\det G_{ij}) \times ( 1 + G^{ij}
A_i A_j ), \nn\\
\det ( G_{ij} + {\cal F}_{ij} ) &=& (\det G_{ij})
\times ( 1 + \frac{1}{2} G^{ij} G^{kl} {\cal F}_{ik} {\cal F}_{jl} ),
\label{4.6}
\end{eqnarray}
where ${\cal F}_{ij} = -{\cal F}_{ji}$.

Eq.(\ref{4.5}) suggests the identification $E^{11} = C_1 + dB$,
in other words, identifying the world-volume scalar with the
coordinate of a compact extra target-space dimension.
Consequently, the Dirac-Born-Infeld action in Eq.(\ref{4.4})
takes the standard form for the induced metric of the M2-brane.
The remaining work is to show that the second term in the right
hand side of Eq.(\ref{4.4}) equals to the expression
for the Wess-Zumino term of the super M2-brane. Indeed, taking
the exterior derivative and using the relation Eq.(\ref{4.3})
we can arrive at the following equation:
\begin{eqnarray}
d ( C_3 + b_2 \wedge dB )
&=& \frac{i}{2} \bar{E} \wedge \gamma_{ab} E \wedge E^b \wedge E^a
+ i \bar{E} \wedge \gamma_{11} \gamma_a E \wedge E^a \wedge E^{11}
\nn\\
&=& \frac{i}{2} \bar{E} \wedge \gamma_{\hat{a}\hat{b}} E
\wedge E^{\hat{b}} \wedge E^{\hat{a}}
\nn\\
&\equiv& d \Omega^{11},
\label{4.7}
\end{eqnarray}
where $\hat{a} \equiv (a, 11)$ denotes 11 dimensional index.
As implied in the above last equation, the left hand side in
Eq.(\ref{4.7}) exactly coincides with the Wess-Zumino term in
the super M2-brane action \cite{Sezgin} except antisymmetric field
which we have neglected from the beginning in this paper.
Accordingly, the
dual action (\ref{4.4}) of the super D2-brane can be written to
\begin{eqnarray}
S_D = - \int_{M_3} d^3 \sigma \sqrt{- \det G^{11}_{ij}}
+ \int_{M_3} \Omega^{11},
\label{4.8}
\end{eqnarray}
where $G^{11}_{ij} = E_i^{\hat{a}} E_j^{\hat{b}} \eta_{\hat{a}\hat{b}}$
and $d \Omega^{11} = \frac{i}{2} \bar{E} \wedge \gamma_{\hat{a}\hat{b}} E
\wedge E^{\hat{b}} \wedge E^{\hat{a}}$. Thus, we have proved that the
super D2-brane action in type IIA supergravity background is transformed to
the super M2-brane action with a circular compactified 11th dimension
in eleven dimensional supergravity background through a duality
transformation of the world-volume gauge field as expected from
IIA/M-duality.

It is straightforward to check that we can also
get the relation between the string metric in ten dimensions and
the eleven dimensional metric \cite{Witten2} by introducing a constant
dilaton in the original action in an appropriate way.
To this end, let us begin by the following action with the
dependence of the constant dilaton background:
\begin{eqnarray}
S' = - \int d^3 \sigma e^{-\phi}
\sqrt{- \det ( G_{ij} + {\cal F}_{ij} )}
+ \int e^{-\phi} ( C_3 + C_1 \wedge
{\cal F} ).
\label{4.9}
\end{eqnarray}
The same procedure as before leads to the dual action
\begin{eqnarray}
S_D' = - \int d^3 \sigma e^{-\phi}
\sqrt{- \det G'_{ij}}
+ \int ( e^{-\phi} C_3 + b_2 \wedge dB ),
\label{4.10}
\end{eqnarray}
where
\begin{eqnarray}
G'_{ij} = G_{ij} + ( e^{\phi} \partial_i B + C_i )
( e^{\phi} \partial_j B + C_j ),
\label{4.11}
\end{eqnarray}
which exactly reduces to (\ref{4.4}) and (\ref{4.5}) in the absence
of the dilaton background. Then we can rewrite $S_D'$ in the previous
form $S_D$ of the standard M2-brane action with the obvious rescalings
\begin{eqnarray}
E^{\hat{a}} \rightarrow e^{\frac{1}{3} \phi} E^{\hat{a}}, \
E^{\alpha} \rightarrow e^{\frac{1}{6} \phi} E^{\alpha}.
\label{4.12}
\end{eqnarray}
Comparing $G'_{ij}$ and $G^{11}_{ij}$, one finds the relation
\begin{eqnarray}
G^{11}_{ij} &=& e^{-\frac{2}{3} \phi} G'_{ij} \nn\\
&=& e^{-\frac{2}{3} \phi} G_{ij} + e^{\frac{4}{3} \phi}
( \partial_i B + e^{-\phi} C_i )
( \partial_j B + e^{-\phi} C_j ).
\label{4.13}
\end{eqnarray}
This equation correctly reproduces the relationship between the
string metric in ten dimensions and the eleven dimensional
metric, in particular, the well-known relation, $R_{11}
= e^{\frac{2}{3} \phi}$ \cite{Witten2} read from the coefficient
in front of $(\partial B)^2$ where $R_{11}$ is the radius of
the compactified 11th dimension on a circle.

\section{The super D3-brane}

In this section let us show the self-duality of the super
D3-brane action in the general type IIB supergravity
background in two ways. The first is the semi-classical way
and the second is the exact one without resort to any
semi-classical approximation.

 From Eqs.(1) and (2), the super D3-brane action in the
Einstein metric becomes
\begin{eqnarray}
S &=& S_{DBI} + S_{WZ}, \nonumber \\
S_{DBI} &=& - \int_{M_4}d^4\sigma \sqrt{-\det(G_{ij}+ e^{-\phi/2}
F_{ij}-b_{2ij})
}, \nonumber \\
S_{WZ} &=& \int_{M_4 =\partial M_5}(C_4 + C_2\wedge (e^{-\phi/2}F-
b_2) + \frac{1}{2}C_0 F\wedge F) \nonumber \\
&=& \int_{M_5}I_5,
\label{5.1}
\end{eqnarray}
where we have explicitly written down the dependence of the dilaton 
field.
And the constraints (12) on the field strengths are given by
\begin{eqnarray}
H_{(3)} &= & db_2 = i\bar{E}\wedge\hat{E}\wedge{\cal K}E,
\nonumber \\
R_{(5)} &=& \frac{i}{6}\bar{E}\wedge \gamma_{abc}{\cal E}E\wedge
E^c\wedge E^b\wedge E^a, \nonumber \\
R_{(3)} &=& -i\bar{E}\wedge\hat{E}\wedge{\cal I}E.
\label{5.2}
\end{eqnarray}
{}From these equations and the definitions (3) and (9), $C_4$ and
$C_2$ are determined by the conditions
\begin{eqnarray}
R_{(5)} &=& dC_4 - db_2\wedge C_2 =
\frac{i}{6}\bar{E}\wedge \gamma_{abc}{\cal E}E\wedge
E^c\wedge E^b\wedge E^a,
\nonumber \\
R_{(3)} &=& dC_2 = -i\bar{E}\wedge\hat{E}\wedge{\cal I}E.
\label{5.3}
\end{eqnarray}

\subsection{The semi-classical self-duality}

In this subsection we show that the super D3-brane action in the
general type IIB supergravity background is semiclassically
self-dual. We first consider the case of vanishing dilaton and
axion. Adding a Lagrangian multiplier term
\begin{eqnarray}
S_{\tilde{H}} = \int_{M_4}d^4\sigma\frac{1}{2}\tilde{H}^{ij}
(F_{ij}-2\partial_i A_j ),
\label{5.4}
\end{eqnarray}
to the above action (\ref{5.1}), the equation of motion for
$A_i$ can be solved by $\tilde{H}^{ij} = \epsilon^{ijkl}\partial
_k B_l$ with a dual vector potential $B_i$. Then after substituting
this solution into the action and solving the equation of motion for
$F_{ij}$, we arrive at the dual action $S_D$ of (41)
\begin{eqnarray}
S_D &=& -\int_{M_4}\sqrt{-\det(G_{ij} +\tilde{F}_{ij}+C_{2ij})}
+ \int_{M_4}\Omega_D, \nonumber \\
\Omega_D &=& C_4 -b_2\wedge C_2 + b_2\wedge(\tilde{F}+C_2),
\label{5.5}
\end{eqnarray}
where $\tilde{F}=dB$.

Next let us perform the following $SO(2)$ rotation of the spinor
coordinate $\theta$
\begin{eqnarray}
\theta^{\prime} = \frac{1}{\sqrt{2}}(1 +{\cal E})\theta, \
\bar{\theta}^{\prime}= \bar{\theta}\frac{1}{\sqrt{2}}(1 -{\cal E}),
\label{5.6}
\end{eqnarray}
then the spinor components of the vielbeins rotate in the same way as
explained at the end of Section 3.
Under these $SO(2)$ rotations of vielbeins it is shown that from 
(42) and (43)  the $b_2, C_2$ and $C_4$
transform as follows;
\begin{eqnarray}
b_2^{\prime} = -C_2, \ C_2^{\prime} = b_2, \ C_4^{\prime} = C_4 - b_2
\wedge C_2.
\label{5.7}
\end{eqnarray}
Then the dual action can be written in terms of transformed
fields as
\begin{eqnarray}
S_D &=& -\int_{M_4}\sqrt{-\det(G_{ij}
+\tilde{F}_{ij}-b_{2ij}^{\prime})}
+ \int_{M_4}(C_4^{\prime} +C_2^{\prime}\wedge
(\tilde{F}-b_2^{\prime})).
\label{5.8}
\end{eqnarray}
The resulting action is completely the same  form as the original action
(\ref{5.1}) from which we have started. Thus we have established the
semi-classical self-duality of the super D3-brane action in the
generic type IIB supergravity background.

 It is a straightforward task to introduce the dilaton and
axion fields and establish the semi-classical $SL(2,R)$ self-duality
of the  action. In this case, the $SO(2)$ transformation rules of the
spinor coordinates
$\theta$ , 2- and 4-form potentials $b_2, C_2$ and $C_4$, and the 
dilaton and axion fields $\tau = C_0 +i e^{-\phi}$ are given by
\begin{eqnarray}
\theta^{\prime} = \frac{1}{\sqrt{2(1+e^{2\phi}C_0^2 - e^{\phi}C_0)}}
\left[\sqrt{1+e^{2\phi}C_0^2} - e^{\phi}C_0 + {\cal E}\right]\theta,
\label{5.9}
\end{eqnarray}
\begin{eqnarray}
b_2^{\prime}&=&\frac{1}{\sqrt{1+e^{2\phi}C_0^2}}(-C_2 - e^{\phi}C_0
b_2), \nonumber \\
C_2^{\prime}&=&\frac{1}{\sqrt{1+e^{2\phi}C_0^2}}(- e^{\phi}C_0 C_2
+b_2), \nonumber \\
C_4^{\prime}&=&C_4 - \frac{1}{1+e^{2\phi}C_0^2}C_2\wedge b_2
+\frac{e^{\phi}C_0}{2(1+e^{2\phi}C_0^2)}(C_2\wedge C_2-b_2\wedge
b_2),
\label{5.10}
\end{eqnarray}
and
\begin{eqnarray}
\tau^{\prime} = -\frac{1}{\tau}.
\label{5.11}
\end{eqnarray}
Combining this transformation with the symmetry under a constant shift
of $C_0$ at the classical level, one deduce the $SL(2,R)$ self-duality of
the super D3-brane action.

\subsection{The exact self-duality}

In this subsection we show that the super D3-brane action in the
type IIB supergravity background
satisfies the Gaillard and Zumino (GZ) self-duality condition, thereby
establishing its exact self-duality without resort to any
semiclassical  approximation.

First let us review the GZ duality condition and its some
properties. Given a generic Lagrangian density ${\cal L}(F_{\mu\nu},
g_{\mu\nu},\phi^A) =\sqrt{-g}L(F_{\mu\nu},
g_{\mu\nu},\phi^A)$ in four dimensional spacetime
 which contains a U(1) gauge field strength $F_{\mu\nu}$,
gravitational field $g_{\mu\nu}$ and generic matter fields
$\Phi^A$, the constructive relation is given by
\begin{eqnarray}
\tilde{K}_{\mu\nu} \equiv \frac{\partial L}{\partial
F_{\mu\nu}}, \
\frac{\partial F_{\alpha\beta}}{\partial F_{\mu\nu}} \equiv 
\delta_{\alpha}^{\mu}
\delta_{\beta}^{\nu} -\delta_{\alpha}^{\nu}\delta_{\beta}^{\mu},
\label{5.12}
\end{eqnarray}
where the Hodge dual components for the anti-symmetric tensor
$K_{\mu\nu}$ are defined by
\begin{eqnarray}
\tilde{K}_{\mu\nu} \equiv
\frac{1}{2}\eta_{\mu\nu}^{\rho\sigma}K_{\rho\sigma}, \
\tilde{\tilde{K}}_{\mu\nu}=-K_{\mu\nu},
\label{5.13}
\end{eqnarray}
where $\eta_{\mu\nu\rho\sigma}
=\sqrt{-g}\epsilon_{\mu\nu\rho\sigma}$,
$\epsilon^{0123}=1$ and the signature of $g_{\mu\nu}$ is $(-,+,+,+)$.

If one defines the infinitesimal $SO(2)$ duality transformation by
\begin{eqnarray}
&&\delta F_{\mu\nu} =\lambda K_{\mu\nu},\ \delta K_{\mu\nu} = -\lambda
F_{\mu\nu} , \nonumber \\
&&\delta\Phi^A = \xi^A(\Phi),
\label{5.14}
\end{eqnarray}
then the consistency of the constructive relation (\ref{5.12}) and the
invariance of the field equations under this $SO(2)$ duality
transformation require the following condition:
\begin{eqnarray}
\frac{\lambda}{4}(F_{\mu\nu}\tilde{F}^{\mu\nu}+
K_{\mu\nu}\tilde{K}^{\mu\nu})
+\delta_{\Phi}{\cal L} =0,
\label{5.15}
\end{eqnarray}
and the invariance of the energy-momentum tensor requires
$\delta g_{\mu\nu}=0$. We call the condition (\ref{5.15}) the GZ
self-duality condition \cite{GZ1, GZ2}.

It has been known that the $SO(2)$ duality is lifted to the
$SL(2,R)$ duality by introducing a dilaton $\phi$ and an axion
$\chi$ \cite{GR}.
Moreover in \cite{IIK1} it has also been shown explicitly that the
GZ condition (\ref{5.15}) is actually the necessary and sufficient 
condition in order  that one can define the off-shell (non-local) 
duality transformation for the $U(1)$ gauge potential itself under 
which the action is invariant.
Therefore if one can show for an action to satisfy the condition
(\ref{5.15}) under the transformation (\ref{5.14})
with suitable transformation
rule for matter fields, then one establishes the exact self-duality
of the theory described by this action without resort to any
semiclassical approximation. In \cite{IIK2, Kimura}
it has been shown that
the super D3-brane actions on the flat and $AdS_5\times S^5$
background indeed satisfy the GZ self-duality condition.

Now let us show that the super D3-brane action (\ref{5.1}) in the general
type IIB supergravity background satisfies the GZ duality condition
under the following $SO(2)$ duality transformation:
\begin{eqnarray}
&&\delta F_{ij} =\lambda K_{ij}, \ \delta K_{ij} = -\lambda
F_{ij} , \nonumber \\
&&\delta\theta = -\frac{\lambda}{2}{\cal E}\theta, \
\delta\bar{\theta} =\frac{\lambda}{2}\bar{\theta}{\cal E}, \
\delta X =0.
\label{5.16}
\end{eqnarray}
The $N=2$ spinor coordinates transform as an $SO(2)$ doublet.
As we already noted that the spinor components of the vielbeins transform
as the same way as $\theta$ under the $SO(2)$ transformation;
\begin{eqnarray}
\delta E = -\frac{\lambda}{2}{\cal E}E, \
\delta\bar{E} =\frac{\lambda}{2}\bar{E}{\cal E}.
\label{5.17}
\end{eqnarray}
Therefore using Eqs.(\ref{5.2}) and (\ref{5.3}) we obtain the
$SO(2)$ transformation rule of the 2- and 4-form potentials as follows;
\begin{eqnarray}
\delta b_2 &=& \lambda C_2 , \ \delta C_2 = -\lambda b_2, \nonumber \\
\delta C_4 &=& \frac{\lambda}{2}(C_2\wedge C_2 - b_2\wedge b_2).
\label{5.18}
\end{eqnarray}

Now let us first prove the self-duality of the super
D3-brane action (\ref{5.1}) with vanishing dilaton and axion fields. 
First
let us calculate
$\frac{\lambda}{4}(F_{ij}\tilde{F}^{ij}+
K_{ij}\tilde{K}^{ij})$. From the constructive relation (\ref{5.12}) and
the action (\ref{5.1}) with vanishing $\phi$ and $C_0$ we obtain
\begin{eqnarray}
\tilde{K}^{ij} &=&  \frac{\partial L}{\partial F_{ij}}
\nonumber \\
&=& \frac{\sqrt{-\det G_{ij}}}{\sqrt{-\det(G_{ij}+
{\cal F}_{ij})}}(-{\cal F}^{ij}+{\cal T}\tilde{{\cal
F}}^{ij}) + \tilde{C}_2^{ij},
\label{5.19}
\end{eqnarray}
where we have used the determinant formula for the
four-by-four matrix:
\begin{eqnarray}
\det(G_{ij}+{\cal F}_{ij}) = \det G_{ij}(1+\frac{1}{2}
{\cal F}_{ij}{\cal F}^{ij} - {\cal T}^2), \
{\cal T}\equiv \frac{1}{4}{\cal F}_{ij}\tilde{{\cal F}}^{ij}.
\label{5.20}
\end{eqnarray}
Taking the Hodge dual of (\ref{5.19}), we find
\begin{eqnarray}
K_{ij}=-\frac{1}{2} \eta_{ijkl} \tilde{K}^{kl}
= \frac{\sqrt{-\det G_{ij}}}{\sqrt{-\det(G_{ij}+
{\cal F}_{ij})}}(\tilde{{\cal F}}_{ij}+{\cal T}{\cal
F}_{ij}) + C_{2ij}.
\label{5.21}
\end{eqnarray}
Then we obtain
\begin{eqnarray}
\frac{\lambda}{4}(F_{ij}\tilde{F}^{ij}+
K_{ij}\tilde{K}^{ij})
&=&\frac{\lambda}{4}(2b_{2ij}\tilde{F}^{ij} +2C_{2ij}\tilde{K}^{ij}
-C_{2ij}\tilde{C}_2^{ij} -b_{2ij}\tilde{b}_2^{ij})  \nonumber \\
&=& \frac{\lambda}{4}(4b_2\wedge  F + 4C_2\wedge  K -2C_2\wedge  
C_2 - 2b_2\wedge  b_2).
\label{5.22}
\end{eqnarray}

Next let us calculate $\delta_{\theta}L$. In the language of
differential forms,
\begin{eqnarray}
\delta_{\theta}L&=&\frac{\partial L}{\partial
{\cal F}}\wedge \delta (- b_2) + {\cal F}\wedge \delta C_2 
+\delta C_4 \nonumber \\
&=&\lambda[- K\wedge C_2 - (F - b_2)\wedge b_2
+\frac{1}{2}(C_2\wedge  C_2 - b_2\wedge  b_2)] \nonumber \\
&=&\lambda[-K\wedge C_2 - F\wedge b_2
+\frac{1}{2}(C_2\wedge C_2 + b_2\wedge b_2)].
\label{5.23}
\end{eqnarray}

It is clearly seen that the right-hand sides of (\ref{5.22}) and (\ref{5.23})
exactly cancel with each other and the GZ-duality condition is
indeed satisfied.
As we have proved the invariance of the action under the
infinitesimal $SO(2)$ duality transformation, the action is
also invariant under the finite $SO(2)$ duality transformation.

Now let us discuss the case with  constant non-vanishing dilaton and
axion background. The action is given by (\ref{5.1}). In this case the
$SO(2)$ self-duality is lifted to the $SL(2,R)$ self-duality \cite{GR}.
Let us write the Lagrangian (\ref{5.1}) as
\begin{eqnarray}
\hat{L}(G,F,\theta,\phi,C_0 )=L(G, e^{-\phi/2}F, \theta)
+\frac{1}{4}C_0 F\tilde{F},
\label{5.24}
\end{eqnarray}
where $L(G, F, \theta)$ is the Lagrangian density without dilaton
and axion which satisfies the $SO(2)$ self-duality.
Then if one
define
$\hat{F}= e^{-\phi/2}F$ and
$\hat{K}$ by taking the dual of $\frac{\partial
L(G,\hat{F},\theta)}{\partial
\hat{F}}$, the background dependence is absorbed in the rescaled
variables $(\hat{K},\hat{F})$. These are related with the background
dependent $(K,F)$ by
\begin{eqnarray}
\left(
    \begin{array}{c}
K \\
F
\end{array}
\right)
&=& V
\left(
    \begin{array}{c}
\hat{K} \\
\hat{F}
\end{array}
\right), \nn\\
V &=& e^{\frac{\phi}{2}}
\pmatrix{
e^{-\phi} & C_0 \cr
0 & 1 \cr}.
\label{5.25}
\end{eqnarray}
Here $V$ is a non-linear realization of $SL(2,R)/SO(2)$ transforming
as
\begin{eqnarray}
V \longrightarrow V^{\prime}=\Lambda V O(\Lambda)^{-1}.
\label{5.26}
\end{eqnarray}
Here $\Lambda$ is a global $SL(2,R)$ matrix
\begin{eqnarray}
\Lambda =
\pmatrix{
 a & b \cr
 c & d \cr},
\label{5.27}
\end{eqnarray}
where $a$, $b$, $c$, $d$ are real numbers satisfying $ad-bc=1$,
and $O(\Lambda)$ is an $SO(2)$ transformation
\begin{eqnarray}
O(\Lambda)^{-1}=
\pmatrix{
\cos\lambda & \sin\lambda  \cr
-\sin\lambda & \cos\lambda  \cr}.
\label{5.28}
\end{eqnarray}
The condition that the form of $V$ (\ref{5.26}) is unchanged under the
transformation (\ref{5.27}) determines the $SO(2)$ rotation angle $\lambda$
and the transformation rule of the background fields $\phi$ and
$C_0$;
\begin{eqnarray}
\tan\lambda=\frac{c  e^{-\phi}}{c C_0 +d},
\label{5.29}
\end{eqnarray}
and
\begin{eqnarray}
\tau \rightarrow \tau^{\prime}=\frac{a\tau +b}{c\tau +d},
\label{5.30}
\end{eqnarray}
where $\tau \equiv C_0 + ie^{-\phi}$.

These results show that if the original Lagrangian $L(G,F,\theta)$ is
invariant under the $SO(2)$ duality transformation the extended
Lagarangian $\hat{L}(G,F,\theta,\phi,C_0 )$
with a dilaton and an axion fields is invariant under the
$SL(2,R)$ duality transformation of $(K, F)$ and $\tau\equiv C_0+
ie^{-\phi}$ and $SO(2)$ rotation of $N=2$ spinor with rotation angle
$\lambda$ given by (\ref{5.29}).

\section{The super D4-brane}
In this section let us start with the super D4-brane action and
perform a duality transformation of the world-volume gauge field
to reach the action obtained by the double-dimensional reduction
of the super M5-brane \cite{Aganagic3, Mario}. The method we
consider is similar to that adopted in Section 4, so we shall
follow a similar path of arguments as in the super D2-brane.
Like the super D2-brane, the analysis in this section is
purely $\it{classical}$.

This time, the super D4-brane action with a Lagrange multiplier
term in the string metric becomes
\begin{eqnarray}
S &=& S_{DBI} + S_{WZ} + S_{\tilde{H}}, \nn\\
S_{DBI} &=& - \int_{M_5} d^5 \sigma
\sqrt{- \det ( G_{ij} + {\cal F}_{ij} )}, \nn\\
S_{WZ} &=& \int_{M_5 = \partial M_6} ( C_5 +
C_3 \wedge {\cal F} + \frac{1}{2} C_1 \wedge
{\cal F} \wedge {\cal F}) = \int_{M_6} I_6, \nn\\
S_{\tilde{H}} &=& \int_{M_5} d^5 \sigma \frac{1}{2}
\tilde{H}^{ij} ( F_{ij} - 2 \partial_i A_j ).
\label{6.1}
\end{eqnarray}
And the constraints (\ref{2.11}) on the field strengths are given by
\begin{eqnarray}
H_3 &=& db_2 = i \bar{E} \wedge \gamma_{11} \hat{E} \wedge E, \nn\\
R_{(6)} &=& \frac{i}{24} \bar{E} \wedge \gamma_{abcd} \gamma_{11} E
\wedge E^d \wedge E^c \wedge E^b \wedge E^a, \nn\\
R_{(4)} &=& \frac{i}{2} \bar{E} \wedge \gamma_{ab} E \wedge
E^b \wedge E^a, \nn\\
R_{(2)} &=& i \bar{E} \wedge \gamma_{11} E.
\label{6.2}
\end{eqnarray}
In this case, $C_5$, $C_3$ and $C_1$ are determined by the conditions
\begin{eqnarray}
R_{(6)} &=& dC_5 + db_2 \wedge C_3 =
\frac{i}{24} \bar{E} \wedge \gamma_{abcd} \gamma_{11} E
\wedge E^d \wedge E^c \wedge E^b \wedge E^a, \nn\\
R_{(4)} &=& dC_3 + db_2 \wedge C_1 =
\frac{i}{2} \bar{E} \wedge \gamma_{ab} E \wedge E^b \wedge E^a, \nn\\
R_{(2)} &=& dC_1 = i \bar{E} \wedge \gamma_{11} E.
\label{6.3}
\end{eqnarray}

As in the case of the super D2-brane, we take the variation with
respect to $A_i$, which gives rise to the solution $\tilde{H}^{ij}
= \frac{1}{6} \epsilon^{ijklm} H_{klm}$ with $H = d B$ with $B$ being
a second-rank tensor superfield \footnote{We apologize for using
often the same alphabet $H$ to express different quantities.
Here $H$ just means the field
strength of the newly introduced tensor field $B$.}.
After substituting this solution into the action, we obtain the action
$S = S_1 + S_2$ where $S_1$ and $S_2$ are defined as
\begin{eqnarray}
S_1 &=& - \int_{M_5} d^5 \sigma
\sqrt{- \det ( G_{ij} + {\cal F}_{ij} )}
+ \int_{M_5} ( {\cal H} \wedge {\cal F} + \frac{1}{2} C_1 \wedge
{\cal F} \wedge {\cal F}), \nn\\
S_2 &=& \int_{M_5} ( C_5 + H \wedge b_2 ),
\label{6.4}
\end{eqnarray}
with ${\cal H} = H + C_3$. The duality transformation amounts to
solving the equation of motion for $F_{ij}$ in order to rewrite the action
in terms of $B$ (or its field strength $H$) instead of $F_{ij}$.
Since $S_2$ does not contain $F$, this part of the action is invariant
under the duality transformation. Therefore, one has only to concentrate
on $S_1$. Following the formula in ref.\cite{Aganagic2}, it is
straightforward to derive the dual action $S_D = S_{D1} + S_2$
where $S_{D1}$ is given by
\begin{eqnarray}
S_{D1} &=& - \int_{M_5} d^5 \sigma
\left[ \sqrt{-G} \sqrt{1 + z_1 + \frac{z_1^2}{2}
- z_2} - \frac{1}{8(1 + C_1^2)} \epsilon_{ijklm}
C^i \tilde{{\cal H}}^{jk} \tilde{{\cal H}}^{lm} \right],
\label{6.5}
\end{eqnarray}
where
\begin{eqnarray}
z_1 &=& \frac{1}{2(-G)(1 + C_1^2)} tr (\tilde{G}\tilde{{\cal H}}
\tilde{G}\tilde{{\cal H}}), \nn\\
z_2 &=& \frac{1}{4(-G)^2 (1 + C_1^2)^2} tr (\tilde{G}\tilde{{\cal H}}
\tilde{G}\tilde{{\cal H}}\tilde{G}\tilde{{\cal H}}
\tilde{G}\tilde{{\cal H}}), \nn\\
G &=& \det{G_{ij}}, \nn\\
\tilde{G}_{ij} &=& G_{ij} +C_i C_j, \nn\\
\tilde{{\cal H}}^{ij} &=& \frac{1}{6} \epsilon^{ijklm} {\cal H}_{klm}.
\label{6.6}
\end{eqnarray}

Now let us consider the Wess-Zumino action $S_2$. The conditions
(\ref{6.3}) yield the equation
\begin{eqnarray}
d (C_5 + H \wedge b_2) = \frac{i}{24} \bar{E} \wedge \gamma_{abcd}
\gamma_{11} E \wedge E^d \wedge E^c \wedge E^b \wedge E^a
- i \bar{E} \wedge \gamma_{11} \hat{E} \wedge {\cal H}.
\label{6.7}
\end{eqnarray}
As a result,  we have the dual action of the super D4-brane in
type IIA supergravity background
\begin{eqnarray}
S_D = - \int_{M_5} d^5 \sigma
\left[ \sqrt{-G} \sqrt{1 + z_1 + \frac{z_1^2}{2}
- z_2} - \frac{1}{8(1 + C_1^2)} \epsilon_{ijklm}
C^i \tilde{{\cal H}}^{jk} \tilde{{\cal H}}^{lm} \right]
+ \int_{M_5} \Omega_D,
\label{6.8}
\end{eqnarray}
where $d \Omega_D = \frac{i}{24} \bar{E} \wedge \gamma_{abcd}
\gamma_{11} E \wedge E^d \wedge E^c \wedge E^b \wedge E^a
- i \bar{E} \wedge \gamma_{11} \hat{E} \wedge {\cal H}$.
This dual action of the super D4-brane is identical to the action
which is obtained by the double-dimensional reduction of the
super M5-brane \cite{Aganagic3}.
(In the above, we have neglected the dilaton field, but
as in the other super D-branes it is easy to include a
constant dilaton background $(p = 1, 2, 3)$ in the present
formulation, from which a more manifest correspondence
of the double dimensional reduction would be obtained.)
Hence we have shown that the
double-dimensional reduction of the super M5-brane action
coincide with the dual super D4-brane action in type IIA supergravity
background as suggested by the duality between M-theory and IIA
superstring theory.

\section{Discussions}
In this paper, we have studied the properties of the duality
transformation of super Dp-brane actions ($p = 1, 2, 3,4$)
in type II on-shell supergravity background. In each case,
the obtained results agreed with the corresponding results
in a flat background. Thus we have succeeded in showing
that various duality symmetries in the super D-brane actions
are independent of the background geometry.

In the last section in a paper \cite{Aganagic2}, it is stated
that "...... For the most part, our analysis has been classical
and limited to flat backgrounds. The results should not depend
on these restrictions, however." In this paper, we have removed
such restrictions completely for the super D1-brane and D3-brane.
On the other hand, for the super D2-brane and D4-brane we have
removed the restriction of 'flat background', but we have
presented only the classical analysis. This restriction should
be also removed in future. Concerning this problem, there may
be a different opinion. Namely, since the Dp-brane actions
with $p > 1$ are in essence unrenormalizable, these actions
might describe the low energy effective theory of underlying
renormalizable theory so that the quantum-mechanical analysis
is too much demanding. This problem still deserves further
investigation.

The present study may also shed some light on symmetries of the
underlying fundamental theory where it is widely believed that
the $SL(2,Z)$ duality found in the super D-string survives as
an exact symmetry of the underlying theory \cite{Hull, Witten2}.
So far, symmetries have given us a useful guiding principle
for establishing a theory in theoretical physics.
Maybe, one of the most challenging studies
in future would be to promote this global
discrete symmetry to the local
gauge symmetry, from which we could draw some very powerful general
conclusions about the relation between the strong coupling phase
and the weak coupling phase and compactified dimensions as well as
the implications for physical four-dimensional spacetime.

Moreover, in the case of type IIB background
we have spelled out the problem of the $SO(2)$
rotation of the $N=2$ spinor coordinates. Our proof utilizes
only an invariance of the constraints and the boundary condition
in the flat background limit so that it can be applied to
the other situations in a straightforward way.

Even if we have limited ourselves to the case of the constant
(or vanishing) dilaton and the vanishing antisymmetric fields,
it may be possible to generalize the analysis adopted in this
paper to a more general situation.
But we should be reminded that even in the case of a flat
background the analysis of dualities is restricted to
be only the constant dilaton field and the vanishing
antisymmetric tensor fields \cite{Aganagic1}.
Perhaps, particularly in the case of type IIB branes,
provided that we would like to consider the non-constant
dilaton, we may have to deal with the non-constant axion
as well on an equal footing since they together magically
parametrize the coset space $SL(2,R)/SO(2)$.

\vs 1
\begin{flushleft}
{\bf Acknowledgement}
\end{flushleft}
The work of I.O. was supported in part by Grant-Aid for Scientific
Research from Ministry of Education, Science and Culture No.09740212.

\vs 1

\end{document}